\documentclass[12pt, a4paper]{iopart}

\usepackage{graphicx}
\usepackage{epsf}
\usepackage{hyperref}
\hypersetup{colorlinks=true,%
            urlcolor=black,%
            linkcolor=black,%
            citecolor=black}

\begin{document}
\title{Academic performance \& student engagement in level~1 physics undergraduates}%
\author{MM~Casey and S~McVitie}
\date{Draft: \today}
\address{Dept. of Physics \& Astronomy, University of Glasgow, GLASGOW G12 8QQ, UK}
\ead{m.casey@physics.gla.ac.uk}
\begin{abstract}
At the beginning of academic year 2007-08, staff in the Department of Physics \& Astronomy at the University of Glasgow started to implement a number of substantial changes to the administration of the level~1 physics undergraduate class. The main aims were to improve the academic performance and progression statistics. With this in mind, a comprehensive system of learning support was introduced, the main remit being the provision of an improved personal contact and academic monitoring and support strategy for {\sl all} students at level~1.  The effects of low engagement with compulsory continuous assessment components had already been observed to have a significant effect for students sitting in the middle of the grade curve. Analysis of data from the 2007-08 class showed that even some nominally high-achieving students achieved lowered grades due to the effects of low engagement. Nonetheless, academic and other support measures put in place during 2007-08 played a part in raising the passrate for the level~1 physics class by  approximately~8\% as well as raising the progression rate by approximately~10\%. 
\end{abstract}
\submitto{Eur. J. Phys. 30 (2009) 1153-1161.}

\maketitle

\section{Introduction}
It is generally accepted that the retention and associated completion rates for first year classes are an area of concern for UK universities~\cite{retention}; level~1 physics at the University of Glasgow is no exception. Classes are often large and, as a result, student integration on academic and social levels can be difficult to achieve; from a staff perspective it can also be difficult to track students who disengage from the process of studying. It had already been observed that students attaining lower grades within the level~1 class frequently failed to present themselves for all available assessment tasks. Until this point, however, the effect of incomplete presentation for assessment on the academic performance of all students in the class not been quantified.

In order to address these issues, staff in the Department of Physics \& Astronomy started to implement a number of substantial changes to the administration of the level~1 physics undergraduate class. One of the authors (Casey) was charged with the responsibility of directing learning support and worked closely with the class head (McVitie). 

The purpose of this paper is to present an overview of some of the changes that were made during the first year of the new r{\'e}gime as well as how a greater understanding of the data from 2007-08 has informed practice in 2008-09.

\section{Undergraduate Physics at the University of Glasgow}
In order to understand how best to address the issue of supporting physics undergraduates, one should first appreciate the structure of science degrees at the University of Glasgow.

\subsection{Undergraduate Science Degrees in Scotland}
Under the Scottish schools' educational system~\cite{lts}, students are able to enter university normally after a minimum of twelve years of primary and secondary education. As a result Scottish undergraduate honours science degrees normally take a minimum of four years to complete. In comparison, students in English and Welsh universities are required to complete thirteen years of primary and secondary education before being admitted. In order to compete with universities in Europe, many UK universities now offer extended undergraduate degrees requiring one further year of study.

\subsection{Faculty Entry at the University of Glasgow}
\label{intention}
In common with some other Scottish universities, the University of Glasgow grants students entry to a Faculty rather than an individual course~\cite{ug}. As well as allowing students to study a broad-based curriculum, one benefit of the Faculty Entry system is that it allows students to revise their intended degree subject at the end of their first year. However, it does mean that the population of the level~1 physics class comprises those intending to pursue a degree in physics as well as those taking the class for credit in order to pursue another degree subject. At the start of their first year, typically $\sim50\%$ of level~1 physics students state an intention to pursue physics to at least level~2.

\subsection{Historical Passrates for Level~1 Physics}
\label{passrate-hist}
Historically, level~1 physics has consisted of two separate modules, Physics~1X (P1X) and Physics~1Y (P1Y). These were assessed separately, Physics~1X at the end of the first semester and Physics~1Y at the end of the second semester. Students were required to achieve at least  D-grade in {\it both} modules in order to progress to level~2 physics. Students were also required to achieve satisfactory grades in the level~1 maths classes, taught separately in the Department of Mathematics. 

The relevant statistic for level~1 physics is that fraction of the class attaining D-grade or above by the end of the academic year. On that basis, the average notional `passrates' over a 5-year period from 2002-2007 were 77.2\% for P1X ($\sigma_{P1X} = 1.9$) and 72.4\% for P1Y ($\sigma_{P1Y} = 2.2$). Our goals in introducing the changes described in this paper were to attempt to raise both passrates as well as narrow the gap in academic performance between Physics~1X and Physics~1Y. 

\subsection{Progression from Level~1 to Level~2 Physics}
A first look at the data showed that in 2006-07, a typical year, some~37\% of the level~1 physics class progressed into the level~2 physics class the subsequent year. If one multiplies the passrate stated in Section~\ref{passrate-hist} with the `intention on entry' figure from Section~\ref{intention}, one should {\it expect} a direct progression rate of approximately 37\%. The question is whether or not that figure can be improved.

\section{Learning Support for First Year}
The primary aim in introducing a comprehensive system of learning support was to enhance the department's effective learning environments by identifying `at-risk' students as well as to  provide extra targeted academic support~\cite{casey}.

One aspect of the Scottish educational system is level~1 university students are typically 17-18 years old compared to those in England and Wales who are, on average, 18-19 years old. It has long been recognised that the some of the younger students can find it difficult to cope with the transition from a well-structured secondary education to the more flexible lifestyle at university~\cite{transition}. The assessment data from previous years provided evidence that some students failed level~1 physics not because they were intellectually incapable but because they did not submit required continuous assessment work on time or, indeed, at all. Discussions with students in higher years revealed that the failure to submit required work on time was often due to the student being unaware that a deadline existed or that the work was required, even though these had been communicated on multiple occasions and in many ways. 

\subsection{SMS Early Warning System}
\label{sms}
Taking cues from work carried out at the University of Brighton~\cite{harley} it was decided that the Short Message Service (SMS) of mobile telephones could be used to great effect. To avoid presenting the role of Director of Learning Support in a negative light, it was decided that students should not be required to provide a mobile telephone number, rather that they should be asked to volunteer it. In fact, nearly all students volunteered a mobile telephone number. 

It was decided that the Director of Learning Support would contact students by SMS immediately upon them missing any continuous assessment component of the class. This was completely different to the approach taken in previous years where students were contacted about missing work at the end of the first semester. For students who had missed a number of deadlines, the end of the first semester could be too late for them to do anything constructive and, invariably, they failed the course. It was hoped, therefore, that contact early in the semester might lessen the impact of missed deadlines and help more students succeed.

\subsection{Extra Academic Support}
Traditional methods for extra academic support typically consist entirely of a series of extra revision classes or large-class tutorials. However, revision classes are usually optional and tend to attract the only academically committed who do not necessarily need the extra help~\cite{biggs}. Therefore, it was decided that a more proactive approach was needed. All students would be able to make use of optional drop-in tutorials, run by a few members of staff providing one-to-one tuition and students who appeared to be underachieving would be personally invited to attend. By presenting the nature of this extra academic support as part of an integrated whole-class approach it was hoped that a positive academic outcome would be achieved.

\section{New Initiatives in the Administration of Level~1 Physics in 2007-08}
\subsection{Continuous Assessment}
Continuous assessment has long been known as a means of helping the weaker students improve their overall grade~\cite{dixon} as well as improving learning~\cite{gibbs}. Therefore, as part of the overall restructure of the level~1 physics course, the proportion continuous assessment contributed to the final grade was increased from 20\% to 40\% with the remainder of the assessment being taken from an end-of-course degree exam. Previously, continuous assessment had been attributed entirely to laboratory work. 

In 2007-08, the class head introduced a series of four 20-minute multiple choice (MCQ) workshop-tests for each of Physics~1X and Physics~1Y. These were worth 20\% in total or 5\% per test per module. The MCQ format was chosen because it meant that the tests could be marked automatically, a necessity when the class contains upwards of $\sim$180~students. The proportion of continuous assessment assigned to laboratory work remained the same as in previous years as did the breakdown of individual components within laboratory work. From the point of view of class administration, the increase in the frequency of continuous assessment tasks presented the possibility that student absence without good cause might also increase. It was always seen as important to the staff (if not the students) that attendance should be maximised. 

\subsection{Monitoring Attendance}
Aside from the obvious benefit of attending continuous assessment component in terms of a student's final grade, it is a Departmental requirement that students attain a minimum of 50\% attendance at workshop-tests and, separately, laboratory classes. Students failing to make these minimum criteria would run the risk of being awarded {\it Credit Refused} for the module. The award of {\it Credit Refused} is a serious impediment to progression.

All students at the University of Glasgow are issued with credit-card sized ID cards which have a barcode printed on them. Portable handheld barcode scanners were obtained from a commercial outlet~\cite{barcode} and used to monitor attendance at every lecture, lab and workshop-test. Using this attendance data, students were contacted on every occasion that they missed a compulsory assessment task. The lecture attendance data was used to further inform student attendance profiles. 

\subsection{Contacting Students}
A number of methods of contacting students were used including the use of the mobile telephone SMS (Section~\ref{sms}). Students were also contacted in person, using email and, if necessary, snail-mail. This order of these contact methods represents an escalation of severity in terms of student absence. 
\begin{itemize}
\item With a view to adopting a more personal approach to the subject of class administration, some considerable time and energy was spent by both authors in approaching students personally. This was done during lab-time as well as by making personal appearances after lectures. Face-to-face conversations were found to be the most effective method of communicating with most students. However, for absent students, other methods became necessary. 
\item SMS was found to be very useful near the start of the academic year to contact students who had missed labs or workshop tests due to their confusion of timetabling or a difficulty in adapting to to life at university. In this early transition period it served to get students back on track quickly.  
\item After a successful pilot scheme in 2005-06, the Department had moved all of its undergraduate course administration to the electronic teaching tool Moodle~\cite{moodle}. All course materials and, more importantly, course notices are posted electronically (traditional noticeboards are rarely used in the Department now); students receive all class information either in person at lectures or by emails sent through Moodle to their student email account. By approximately half-way through the first semester, when students had adjusted to reading their email accounts regularly, email was the primary method used for contacting absent students. 
\item Nonetheless, some students failed to respond to repeated attempts to contact them either in person, by SMS or by email. In cases of continual non-attendance, students received personal letters which were copied to the student's Adviser of Studies, the person responsible for overseeing a student's academic progression with the university.
\end{itemize}
The results from these efforts are discussed in section~\ref{results}.

\section{Results and Analysis}
\label{results}
\subsection{Attendance Profiling}
\label{att}
In terms of absenteeism from compulsory continuous assessment components, it was found that the student population could be broken down into three broad categories:
\begin{itemize}
\item Group~1 ($\sim 90\%$ of the population): Most students in this category achieved full attendance. The remainder would miss a single isolated session due to a one-off event that could be attributed to confusion with the timetable (near the start of the academic year), illness or another compassionate absence reason. These students {\it always} responded to attempts to contact them, be it in person, by SMS or by email. The attendance of students in this group was, as a whole, very high. The academic performance of students in this group was primarily in the A-grade and D-grade range, with some achieving E-grades
\item Group~2 ($\sim 5\%$ of the population): Students in this category missed multiple sessions. It was difficult to find such students in person (because their attendance was poor) and their response to SMS or email was sporadic. When sent a letter, they sometimes responded and their attendance sometimes improved for a period before falling off again. The academic performance of students in this group tended to fall in the C-grade to E-grade range. However, some students did manage to achieve A-grades or B-grades even with sporadic attendance. 
\item Group~3 ($\sim 5\%$ of the population): Students in this category rarely responded to any attempt to contact them. When eventually contacted, such students normally cited a catalogue of reasons as to why they had missed sessions but it was rare to be able to obtain from them the required documentary evidence to support such absences. Generally speaking, the attendance of such students did not improve at all. The academic performance of students in this group was typically E-grade and below, with a large proportion being refused credit for the course. 
\end{itemize}
Initially, it had been hoped that all students' attendance could be improved with the application of personal effort. However, it was found that there was a hard core of students (Group~3) who appeared to be unresponsive to the offers of assistance. In fact, in order to manage the workload associated with supporting students more effectively, a policy of `five strikes' was adopted by the second semester of 2007-08. If students did not respond within five consecutive attempts to contact them, effort was diverted to supporting those students who were in good attendance. Although five consecutive `strikes' could be accrued relatively quickly during the semester, this number was found to be an extremely good indicator of a student's membership of Groups~2 or~3; in fact, by the second semester it was dropped to three `strikes'. 

\subsection{Effects of Absenteeism on Academic Performance}
It had already been suspected that the effect of incomplete presentation for continuous assessment tasks was quite large for the weaker students (those who normally achieve D/E grades). However, another observation had been made during the marking of final exam papers -- some students did not answer all the questions in the paper, thus throwing away 7.5\% of their final mark for each question not attempted. 

\begin{figure}[ht!]
\begin{center}
\includegraphics[width=.5\textwidth]{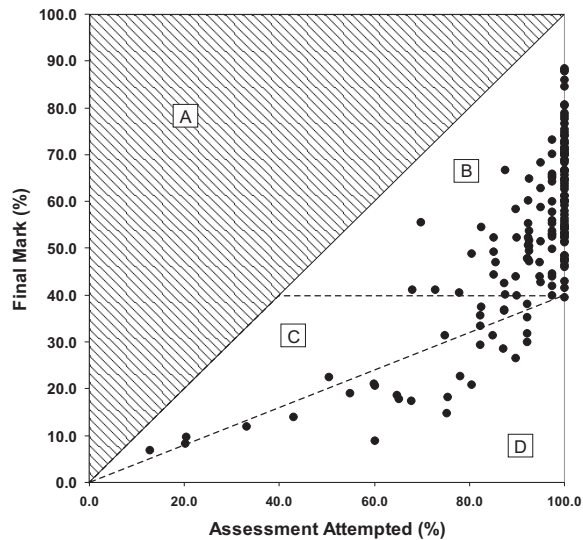}
\caption{Academic performance versus the percentage of assessment attempted for students in Physics~1Y, 2007-08. Triangles A, B, C and D indicate four regions of interest in the data. The maximum mark a student can attain must be equal to or less than the amount of assessment they attempt. Therefore, region~A must remain unpopulated. The D/E-grade boundary for Physics~1Y is set at 40\% and, therefore, regions~B and~C indicate respectively those students who had `passed' and `failed' Physics~1Y. The significance of region~C is that students whose academic performance lies within this region could been placed in region~B (a `pass' grade) had they attempted more assessment tasks. Students whose academic performance placed them in region~D could not have exceeded a 40\% final mark even if they had maximised their attendance. }
\label{DE-gradeslipping}
\end{center}
\end{figure}

As a means of quantifying these effects, firstly a scatter-graph of the percentage of assessment that students presented (x-axis) for against their final percentage mark (y-axis) was made. The plot is shown in Figure~\ref{DE-gradeslipping}.

\begin{figure}[ht!]
\begin{center}
\includegraphics[width=.45\textwidth]{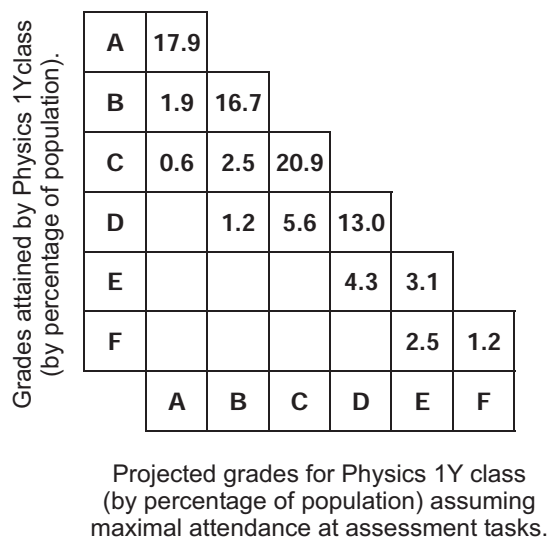}
\caption{Cross-table showing a breakdown of grades in Physics~1Y in 2007-08. Students' actual grades are compared with their projected grades to show the effects of grade-slippage in the class. For example, the third row indicates that 20.9\% of the class attained a C-grade and that this was in line with their projected grade (column); they could not have improved their grade anymore because their attendance was already very high. However, in this same row, 2.5\% of the class attained a C-grade but could have achieved a B-grade with improved attendance. More pertinent, 0.6\% of the class could have improved from a C-grade to an A-grade. NB: The numbers do not add up to 100\% because those students for whom credit was refused in Physics~1Y have been removed from the data.}
\label{mileage}
\end{center}
\end{figure}

To further quantify the effect shown in Figure~\ref{DE-gradeslipping}, a linear scaling was applied to all students' grades to project the grade they could have achieved had they maintained their academic performance and maximsed their attendance. This analysis is shown in Figure~\ref{mileage} for Physics~1Y in 2007-8.

The effect of incomplete assessment is, perhaps predictably, more severe for students in the middle and lower parts of the grade distribution but, as can be seen from the data in Figure~\ref{mileage}, students at all levels slip grades due to this effect. This has a more subtle effect than might at first be imagined. Although the grades a student achieves at level~1 do not contribute to their final degree classification, they do determine entrance (or otherwise) to the advanced honours study programmes.

\subsection{Passrates}
\begin{figure}[ht!]
\begin{center}
\includegraphics[width=.75\textwidth]{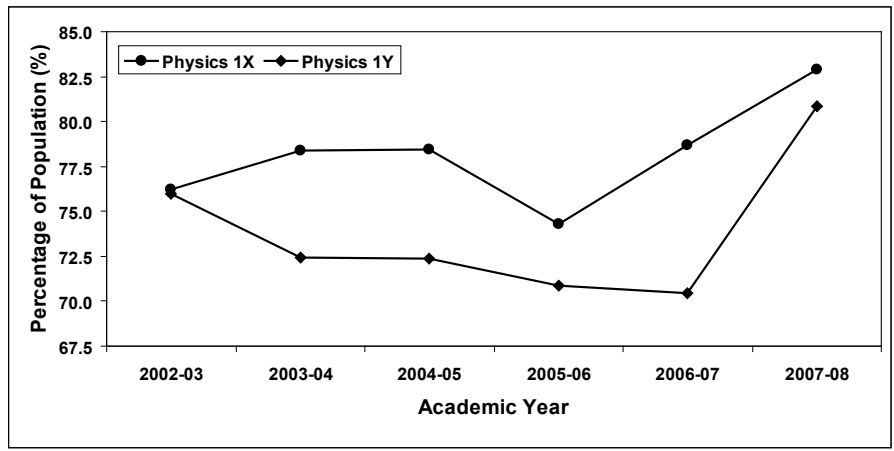}
\caption{Percentage of the class attaining A-D passes after the resit diet in Physics~1X and Physics~1Y by academic year.}
\label{passrate}
\end{center}
\end{figure}

In terms of progression to level~2, students must attain at least a grade-D (40\% or above) in both Physics~1X and Physics~1Y although this can include results from the resit diet of exams. Data from years 2002-03 to 2007-08, shown in Figure~\ref{passrate}, show that both  significant improvements in passrates for both Physics~1X and Physics~1Y and a reduction in the historical gap in attainment between Physics~1X and Physics~1Y had been achieved. Compared to the historical averages, the P1X passrate had increased by 5.7\% (and $3~\sigma_{P1X}$) to 82.9\% while P1Y had increased by 8.5\% (and $4~\sigma_{P1Y}$) to 80.9\%.

\subsection{Drop-in Tutorials: Attendance and Performance}
In 2007-08, optional drop-in one-to-one tutorials were offered for the first time. The aim of the tutorials was partly to address the historical drop in attainment between Physics~1X and Physics~1Y but also to help the weaker students in the class. The tutorials were open to all students in the class, but those falling into at-risk categories (determined by monitoring academic performance on an on-going basis) were sent a personal invitation to attend. Even so, only $\sim$~10\% of the whole class attended any of the drop-in tutorials and these were primarily students in the B-D bands. Even a personal approach was insufficient to encourage attendance in students who either previously or subsequently attained E-grade or less. 

In terms of academic attainment, those students who did attend exhibited an average increase of 2.9\% between their P1X and P1Y marks whilst the class as a whole averaged a drop of 1.6\% between P1X and P1Y. When questioned, students who `opted-in' found the tutorials useful. Therefore, in spite of the low attendance, the drop-in tutorials were deemed to be a success and their provision was expanded in 2008-09.

\subsection{Progression to Level~2}
At the start of 2007-08, 50\% of level~1 students had self-identified an `intention to progress' to level~2 physics. Based on the passrate, a direct progression rate of $\sim40\%$, in keeping with previous years, was expected. In fact, data from the 2008-09 level~2 class showed that the direct progression rate was 47\%, a significant improvement on previous years.

\section{Summary and Future Work}
The main aims in the restructuring of the level~1 physics course appear not only to have been achieved but, on some levels, to have been surpassed. The 2007-08 passrate in both Physics~1X and Physics~1Y increased by approximately~8\%, a move exceeding 3~standard deviations from the historical averages for the classes. More significantly, the direct progression rate from level~1 physics to level~2 physics was increased by approximately~10\%. It is expected that these levels will be maintained in 2008-09.

A number of methods of contacting and supporting students during their first year of undergraduate study were used including personal approaches, SMS, email and snail-mail. The first three methods were found to be very effective in supporting and improving the experience of students who could be considered to be engaged (Group~1 in Section~\ref{att}). There was even some success in encouraging semi-engaged students to improve attendance (Group~2 in Section~\ref{att}. However, another hard-core of non-attending, non-engaged students was identified and it appeared that little could be done to support or improve their experience (Group~3 in Section~\ref{att}). This is an observation made by staff in other Departments within the University of Glasgow as well as further afield; an investigation of the effectiveness of additional support strategies where non-academic issues appear dominant is necessary.

Experience gained in 2007-08 was used to inform practice in 2008-09 and preliminary results from the level~1 and level~2 2008-09 classes are encouraging in terms of improved academic achievement and lower absenteeism. 

Future work will centre around methods to streamline the data capture techniques so that these may be  rolled out to other classes within the Department, notably level~2 physics. In particular, an investigation into understanding academic performance in level~2 physics and the related level~2 to level~3 progression statistics is now being undertaken. A number of longitudinal studies will be made, starting with the 2007-08 level~1 cohort.

\newpage

\mbox{\hspace{0.5cm}}

\end{document}